\documentclass[a4paper,12pt]{article}

\usepackage{amsmath,amssymb,amsfonts}
\usepackage{graphicx}
\usepackage{color}
\makeatletter
\@addtoreset{equation}{section}
\renewcommand{\theequation}{\thesection.\@arabic\c@equation}
\makeatother
\usepackage{hyperref}
\usepackage{cite}
\usepackage{caption}
\definecolor{red}{rgb}{1,0,0}
\definecolor{green}{rgb}{0,1,0}
\definecolor{blue}{rgb}{0,0,1}
\definecolor{darkblue}{rgb}{0,0,0.5}
\definecolor{lightblue}{rgb}{.5,.5,1}
\definecolor{lightgray}{gray}{.87}
\definecolor{Dark}{gray}{.20}
\definecolor{pink}{rgb}{.95,0.82,0.92}
\definecolor{yellow}{rgb}{1,1,0}
\definecolor{lightyellow}{rgb}{1,1,.5}
\definecolor{purple}{rgb}{0.7,0,0.85}
\definecolor{darkgreen}{rgb}{0,0.5,0}
\definecolor{orange}{rgb}{0.8,0.2,0.2}
\def \be {\begin{equation}}
\def \ee {\end{equation}}
\def \bea {\begin{align}}
\def \eea {\end{align}}
\def \nn {\nonumber}

\def \rr {\raise.35ex\hbox{\small $\prime$}\kern-.17em{\mbox{\large $\imath$}}}
\def \del {\partial}
\def \dels {\partial\kern-.5em / \kern.5em}
\def \As {{A\kern-.5em / \kern.5em}}
\def \Ds {D\kern-.7em / \kern.5em}

\def \dag {\dagger}

\def \eps {\epsilon}

\def \lam {\lambda}
\def \Lam {\Lambda}

\def \om {\omega}
\def \Om {\Omega}

\def \th {\theta}

\newcommand{\detail}[1]{}

\newcommand{\hide}[1]{}

\newcommand{\explanation}[1]{}
\newcommand{\noneed}[1]{}

\pdfoutput=1

\begin{document}

\pagestyle{plain}


\begin{titlepage}
\vspace*{-10mm}   
\baselineskip 10pt   
\begin{flushright}   
\begin{tabular}{r} 
\end{tabular}   
\end{flushright}   
\baselineskip 24pt   
\vglue 10mm

\begin{center}

\noindent
\textbf{\LARGE
From Uneventful Horizon to Firewall\\
in $D$-Dimensional Effective Theory
}
\vskip20mm
\baselineskip 20pt

\renewcommand{\thefootnote}{\fnsymbol{footnote}}

{\large
Pei-Ming~Ho 
\footnote{pmho@phys.ntu.edu.tw},
}

\renewcommand{\thefootnote}{\arabic{footnote}}

\vskip5mm

{\it  
Department of Physics and Center for Theoretical Physics, \\
National Taiwan University, Taipei 106, Taiwan,
R.O.C. 
}

\vskip 25mm
\begin{abstract}

Assuming the standard effective-field-theoretic formulation of Hawking radiation,
we show explicitly how a generic effective theory predicts a firewall from an initially uneventful horizon
for a spherically symmetric, 
uncharged black hole in $D$ dimensions for $D \geq 4$.
The firewall is created via higher-derivative interactions
within the scrambling time after the collapsing matter enters the trapping horizon.
This result manifests the trans-Planckian problem of Hawking radiation
and demonstrates the incompatibility between
Hawking radiation and the uneventful horizon.

\end{abstract}
\end{center}

\end{titlepage}

\pagestyle{plain}

\baselineskip 18pt

\setcounter{page}{1}
\setcounter{footnote}{0}
\setcounter{section}{0}


\newpage

\section{Introduction}
\label{introduction}

It is well known that 
an $\mathcal{O}(1)$-correction to the effective theory
\cite{Mathur:2009hf}
(such as a firewall \cite{firewall,firewall-B}) around the horizon
is crucial to resolve the information loss paradox 
\cite{Hawking:1976ra,Mathur:2009hf,Polchinski:2016hrw,Marolf:2017jkr}.
But,
until very recently \cite{Scattering},
there is no known mechanism in the effective theory
to explain what the $\mathcal{O}(1)$-correction is
and how it arises.

More precisely,
information loss paradox disappears
if there is no Hawking radiation.
The crux of the paradox is the conflict between
Hawking radiation and the uneventful horizon.
When (and only when) there is Hawking radiation, 
there must be drama at the horizon.

In the conventional model of black holes,
the horizon is assumed to be in vacuum
for freely falling observers
due to the equivalence principle.
Higher-derivative interactions violate
the equivalence principle
(see e.g. Ref.\cite{Lafrance:1994in}
in the context of classical electrodynamics),
but they are suppressed by negative powers of the cutoff energy scale $\Lambda$.
Surprisingly,
it was shown for 4-dimensional dynamical black holes \cite{Scattering} that,
for a generic low-energy effective theory,
the assumptions needed for the derivation of Hawking radiation \cite{Hawking-Radiation}
also implies the emergence of a firewall
\footnote{
Here, the phrase ``firewall'' does not refer to a divergent energy flux,
but only a high-energy flux of a scale comparable to $\Lambda$
with respect to the comoving frame of the collapsing matter.
}.
In this paper,
we extend this conclusion to $D$ dimensions for $D \geq 4$.

The origin of this connection between Hawking radiation and the firewall
is the trans-Planckian problem \cite{trans-Planckian-1}.
Hawking particles are originated from quantum modes
with trans-Planckian frequencies at the horizon.
We find that,
after the collapsing matter has entered the uneventful horizon,
the transition amplitude for the creation of these trans-Planckian modes
through higher-derivative interactions becomes large at the horizon
within the scrambling time.

If one wishes to avoid the firewall,
one has to claim that 
the effective theory breaks down for these trans-Planckian modes.
But this also implies that Hawking radiation is not a reliable prediction.
A possibility is that Hawking radiation is turned off
within the scrambling time
and the black hole becomes classical.
Another possibility is that the firewall appears,
and the horizon is no longer uneventful.
In both scenarios, 
the effective theory is invalid within the scrambling time.

In this work,
we adopt the convention that
$\hbar = c = 1$,
and define the Planck length $\ell_p$
and the Planck mass $M_p = 1/\ell_p$
by $G_N = \ell_p^{D-2}$,
where $G_N$ is the Newton constant in the $D$-dimensional spacetime.
We shall always assume that
the Schwarzschild radius $a$ of the black hole
is much larger than $\ell_p$.

\section{Effective theory in $D$ dimensions}

To investigate the self-consistency of the effective-theoretic description
of the black-hole evaporation,
we start by examining the assumptions about effective theories and Hawking radiation.

For a generic low-energy effective theory with a cutoff energy $\Lambda$,
its Lagrangian is in principle an expansion of all local operators $\{\hat{\cal O}_n\}$
(see e.g. \S 12.3 of Ref.\cite{Weinberg:1995mt}):
\begin{align}
{\cal L} = {\cal L}_0 + \sum_n \frac{\lam_n}{\Lambda^{d_n-D}} \, \hat{\cal O}_n,
\label{L}
\end{align}
where ${\cal L}_0$ is the free-field Lagrangian,
$\lam_n$'s are the dimensionless coupling constants,
and $d_n$'s are the dimensions of the local operators $\hat{\cal O}_n$.

\subsection{Physical States}
\label{states}

The effective theory is only applicable to physical states
of energy scales much lower than the cutoff energy scale $\Lambda$.
But since the energy of a state can be arbitrarily large after a Lorentz boost,
the energy bound should only be applied to Lorentz-invariant quantities.
For instance, 
a state composed of two particles of momenta $p^{(1)}$ and $p^{(2)}$
should not be considered in the effective theory
if $\left|g^{\mu\nu} p^{(1)}_{\mu} p^{(2)}_{\nu}\right| \gg \Lambda^2$,
but the energy $E^{(i)} = p^{(i)}_0$ of either particle 
is allowed to be much larger than $\Lambda$.

In the conventional model of black holes
(see e.g. Refs.\cite{Brout:1995rd,Frolov:1998wf} for reviews),
there are three states 
that must be included in the effective theory
if it predicts the usual Hawking radiation \cite{Scattering}.

The first state that must be included in the effective theory is 
the vacuum state $|0\rangle$ for freely falling observers.

We denote by $U$ the outgoing light-cone coordinate 
which can be identified with the Minkowski null coordinate of the infinite past,
and by $u$ the outgoing light-cone coordinate
which can be identified with the Minkowski retarded time coordinate 
at large distances.

The vacuum $|0\rangle$ is annihilated by
the annihilation operator $a_{\om}$ ($\tilde{a}_{\om}$)
associated with the negative-frequency modes $e^{-i\om U}$ ($e^{-i\om V}$),
but not by $c_{\om}$ ($\tilde{c}_{\om}$) associated with $e^{-i\om u}$ ($e^{-i\om v}$).
The operators are related via a Bogoliubov transformation:
\begin{align}
c_{\om} = 
\int_0^{\infty} d\om' \, (A_{\om\om'} a_{\om'} + B_{\om\om'} a_{\om'}^{\dag}).
\label{c=Aa}
\end{align}
The expectation value of the number of Hawking particles
of a given frequency $\om$ for fiducial observers is
\begin{align}
\langle 0|c_{\om}^{\dag}c_{\om}|0\rangle 
= \, \parallel c_{\om}|0\rangle \parallel^2.
\end{align}
For a black hole with the Schwarzschild radius $a$,
if the prediction of Hawking radiation is reliable,
the state $c_{\om}|0\rangle$ for $\om \sim \mathcal{O}(1/a)$
must also be included in the effective theory.
Otherwise,
one cannot be certain about
the spectrum of Hawking radiation.

Apart from $|0\rangle$ and $c_{\om}|0\rangle$ ($\om \sim \mathcal{O}(1/a)$),
the 3rd class of states that should exist in the effective theory is 
the multi-particle states
$a_{\om'_1}^{\dag}\cdots a_{\om'_k}^{\dag}|0\rangle$
and $\tilde{a}_{\om'_1}^{\dag}\cdots \tilde{a}_{\om'_k}^{\dag}|0\rangle$
with sufficiently low energies $\om'_i$.
(We will consider the limit $\om_i \rightarrow 0$.)
We shall now consider compositions of these three classes of states.

Assume that there are three scalar fields $\phi_1$, $\phi_2$, and $\phi_3$
\footnote{
The discussion below will be essentially the same if $\phi_1 = \phi_2 = \phi_3$.
}
in the effective theory.
The Hilbert space is ${\cal H} = {\cal H}_1\otimes{\cal H}_2\otimes{\cal H}_3$,
where ${\cal H}_I$ ($I = 1, 2, 3$) is the Fock space for each field $\phi_I$.
We define the following two states:
\begin{align}
|i\rangle &\equiv |0\rangle\otimes|0\rangle\otimes|0\rangle,
\label{i}
\\
|f\rangle &\equiv \sqrt{2\om} \, c_{\om}|0\rangle
\otimes \sqrt{2\om'} \, \tilde{a}^{\dag}_{\om'}|0\rangle
\otimes \left[\prod_{i=1}^{m} \sqrt{2\om''_i} \, \tilde{a}^{\dag}_{\om''_i}\right]|0\rangle,
\label{f}
\end{align}
where $\om\sim\mathcal{O}(1/a)$,
$\om'$ and $\om''_i$ are arbitrarily close to $0$
so that there is no large invariant energy scale due to $\om'$ or $\om''_i$.
In particular,
$\om'$ and $\om''_i$ are assumed to satisfy
\begin{align}
\left(\frac{dU}{du}\right)^{-1} \om\om' \ll \Lambda^2,
\qquad
\left(\frac{dU}{du}\right)^{-1} \om\om''_i \ll \Lambda^2,
\label{omomp}
\end{align}
where $(dU/du)^{-1}\om$ is the frequency defined with respect to $U$.
(Recall that the frequency parameter $\om$ of $c_{\om}$ 
is defined with respect to $u$.)
From our discussions above,
the states \eqref{i} and \eqref{f} must be included in the effective theory.
We are interested in the transition amplitude
of the process $|i\rangle \rightarrow |f\rangle$.

\subsection{Breakdown of Effective Theory}
\label{Breakdown-Criteria}

The derivation of Hawking radiation
assumes that the free field theory is a good approximation.
In particular,
all higher-dimensional (non-renormalizable) operators $\hat{\cal O}_A$ are ignored.
On the other hand, if
\begin{align}
\left| \frac{\lam_A}{\Lambda^{d_A - D}} \int d^D x \, \sqrt{-g} \; \langle f| \hat{\cal O}_A |i\rangle \right|
\gtrsim \mathcal{O}(1)
\label{precond}
\end{align}
for certain higher-dimensional operators $\hat{\cal O}_A$,
the conventional prediction of Hawking radiation
would be significantly modified.

For the static Schwarzschild background,
or any geometry with a time-like Killing vector,
the translation symmetry in time implies energy conservation,
which in turn forbids a transition from the vacuum to a multi-particle state.
Therefore,
the back-reaction of the vacuum energy-momentum tensor
is crucial for the condition \eqref{precond} to hold.

In Ref.\cite{Scattering},
the condition \eqref{precond} has been shown to be satisfied
by infinitely many higher-derivative operators $\hat{\cal O}_n$
for $D = 4$ when the collapsing matter is deep inside the near-horizon region.
In the following,
we extend this calculation to $D \geq 4$.

\section{Near-Horizon Geometry}
\label{Near-Horizon-Geometry}

In this section,
we solve the near-horizon geometry from the semi-classical Einstein equation.
It is an extension of the 4D result obtained in Refs.\cite{Ho:2019pjr,ShortDistance,Scattering}.

Consider the metric for a spherically symmetric dynamical black hole
with the ansatz
\begin{align}
ds^2 &= - C(u, v) du dv + r^2(u, v) d\Omega^{D-2}.
\label{metric}
\end{align}
We shall give the solution for $C(u, v)$ and $r(u, v)$ in the near-horizon region\footnote{
see Refs.\cite{ShortDistance,Scattering}
for a more precise definition of the near-horizon region.
}
to the semi-classical Einstein equation
\begin{align}
G_{\mu\nu} = \kappa \langle T_{\mu\nu} \rangle,
\label{SCEE}
\end{align}
where $\kappa = 8\pi G_N$.

\subsection{Uneventful Horizon}
\label{Uneventful-Horizon}

\noneed{
As a quantum effect, 
the vacuum energy-momentum tensor $\langle T_{\mu\nu} \rangle$
is a first-order term in $\hbar$.
Since $\langle T_{\mu\nu} \rangle$ is multiplied with $\kappa$ in the equation above,
an expansion in $\hbar$ is equivalent to an expansion in $\kappa$
for our convention in which $\hbar = 1$.
}

According to the equivalence principle,
the vacuum energy-momentum tensor $\langle T_{\mu\nu} \rangle$
should not be much larger than the energy scale of $\mathcal{O}(1/a)$
for freely falling observers comoving with the collapsing matter.
That is,
one assumes the condition of ``uneventful horizon''
\cite{Fulling:1977jm,Christensen:1977jc},
\begin{align}
\langle T_{uu}\rangle &\sim \mathcal{O}(C^2/a^D), 
\label{Tuu}
\\
\langle T_{uv}\rangle &\sim \mathcal{O}(C/a^D),
\label{Tuv}
\\
\langle T_{vv}\rangle &\sim \mathcal{O}(1/a^D),
\label{Tvv}
\\
\langle T_{\th\th}\rangle &\sim \mathcal{O}(1/a^{D-2}),
\label{Tthth}
\end{align}
in the conventional model of black holes.
This assumption has been adopted by most of the works on black holes.
(See, e.g. Refs.
\cite{Davies:1976ei,Fulling:1977jm,Christensen:1977jc,Brout:1995rd,Frolov:1998wf,Parentani:1994ij,Fabbri:2005zn,Fabbri:2005nt,Barcelo:2007yk}.)

\begin{figure}
\center
\includegraphics[scale=0.5,bb=0 0 500 250]{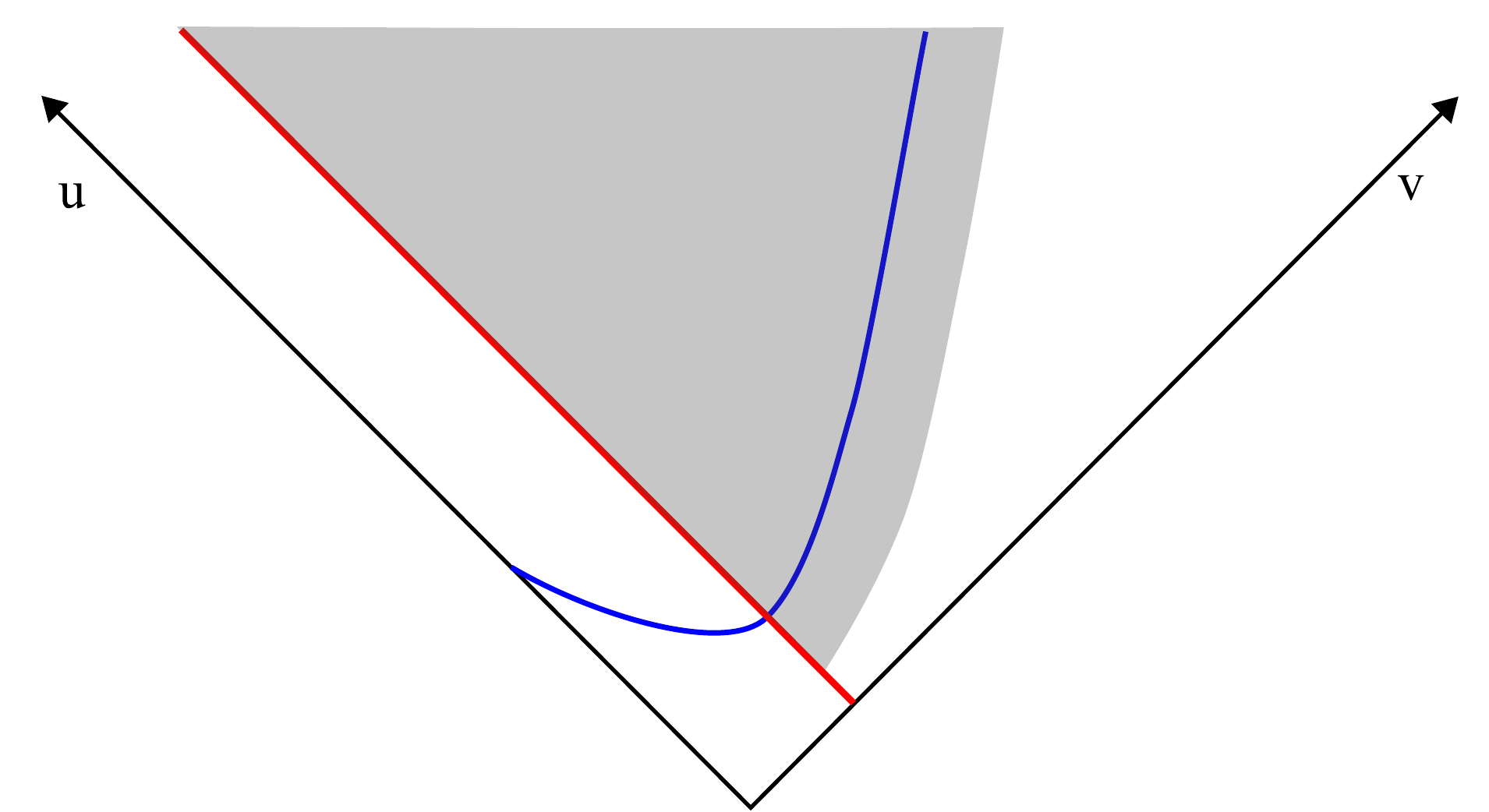}
\caption{\small
The solid blue curve is the trapping horizon,
the red straight line the surface of the collapsing matter
nearly at the speed of light,
and the shaded area the near-horizon region.
}
\label{horizon}
\vskip1em
\end{figure}

Since the outgoing energy flux $\langle T_{uu} \rangle$ \eqref{Tuu}
is extremely small in the near-horizon region due to the tiny conformal factor $C$,
the ingoing energy flux \eqref{Tvv} must be negative,
\begin{align}
\langle T_{vv} \rangle < 0,
\label{Tvv<0}
\end{align}
to account for the decrease in the black-hole mass over time.
The outer trapping horizon in vacuum is hence time-like
\cite{Hayward:2005gi,Ho:2019kte}.
See Fig.\ref{horizon}.

\noneed{
wormhole-like neck (local minimum in areal radius),
apparent horizon when dynamical,
Wheeler's bag of gold, 
event horizon,
information loss.
}

\subsection{Geometry for Freely Falling Observers}
\label{Geometry-FFO}

The near-horizon metric for $D\geq 4$ is found in App.\ref{Solution-EE}
as a solution to the semi-classical Einstein equation \eqref{SCEE}
following the same approach adopted in Refs.\cite{ShortDistance,Scattering}.
For the calculation below,
it is sufficient to approximate the solution \eqref{C} and \eqref{r} by
\begin{align}
C(u, v) &\simeq C(u_{\ast}, v_{\ast})
e^{- (D-3) \frac{u - u_{\ast} + v_{\ast} - v}{2 \bar{a}_{\ast}}},
\label{C0}
\\
r(u, v) &\simeq \bar{a}(v),
\label{r0}
\end{align}
where $\bar{a}_{\ast} \equiv \bar{a}(v_{\ast})$.
The reference point $(u_{\ast}, v_{\ast})$ can be
anywhere in the near-horizon region.
The time-dependent Schwarzschild radius $\bar{a}(v)$
decreases with the advanced time $v$ like
\begin{align}
\frac{d\bar{a}(v)}{dv} \simeq - \frac{\sigma \ell_p^{D-2}}{\bar{a}^{D-2}(v)},
\label{dbaradv}
\end{align}
with a parameter $\sigma$ of $\mathcal{O}(1)$
(see eq.\eqref{dadt}).
Throughout the process under consideration,
the change in $v$ is much shorter than $\mathcal{O}(a^{D-1}/\ell_p^{D-2})$,
so that $\bar{a}(v)$ remains the same order of magnitude
which continues to be denote by $\mathcal{O}(a)$.

Eq.\eqref{C0} also applies to the Schwarzschild solution
with the Schwarzschild radius $\bar{a}_{\ast}$.
Its exponential form is a crucial feature of $C(u, v)$ that leads to Hawking radiation.
The lowest-order approximation of
$r(u, v)$ \eqref{r0} is approximately $u$-independent
because everything infalling is nearly frozen in the near-horizon region
from the viewpoint of a distant observer.

We define a new coordinate system $(U, V)$ by
\begin{align}
\frac{dU}{du} &\simeq
c C(u_{\ast}, v_{\ast}) e^{- (D-3) \frac{(u-u_{\ast})}{2\bar{a}_{\ast}}},
\label{dUdu}
\\
\frac{dV}{dv} &\simeq
c^{-1} C(u_{\ast}, v_{\ast}) e^{- (D-3) \frac{(v_{\ast} - v)}{2\bar{a}_{\ast}}}
\label{dVdv}
\end{align}
for an arbitrary constant $c$.
This is the coordinate system suitable for freely falling observers,
and the freedom in choosing $c$ is related to the relative velocity among them.
The metric \eqref{metric} with $C$, $r$ given by eqs.\eqref{C0}, \eqref{r0} becomes
\begin{align}
ds^2 \simeq - dU dV + r^2 d\Omega^{D-2}.
\label{metric-UV}
\end{align}
Here,
$r \simeq \bar{a}(v(V))$ is interpreted as a function of $V$,
where the function $v(V)$ is determined by eq.\eqref{dVdv}.


Note that eq.\eqref{metric-UV} is the flat metric in the $(U, V)$ space.
All local curvature invariants for the metric \eqref{metric-UV}
are of the characteristic length scale of $\mathcal{O}(a)$.
This is compatible with the fact that
$(U, V)$ is the natural coordinate system for freely falling observers.

\section{Effective Theory in Near-Horizon Region}
\label{ET-NHR}


The conventional derivation of Hawking radiation
is based on the free field theory.
The free field equation for a massless scalar field 
in the near-horizon region is $\nabla^2 \phi = 0$.
It is equivalent to
\begin{align}
\del_u\del_v \varphi - \frac{\del_u\del_v\left(r^{(D-2)/2}\right)}{r^{(D-2)/2}}\varphi 
= \del_u\del_v\varphi + \mathcal{O}(C) = 0
\qquad 
\left(\varphi(u, v) \equiv r^{(D-2)/2}\phi\right),
\end{align}
for the $s$-wave modes according to eq.\eqref{r}.
We expand $\varphi$ in Fourier modes in terms of $(U, V)$ as
\begin{align}
\varphi \simeq \int_0^{\infty} \frac{d\Om}{2\pi} \; 
\frac{1}{\sqrt{2\Om}} 
\left(
e^{- i\Om U} a_{\Om} + e^{i\Om U} a^{\dag}_{\Om}
+ e^{- i\Om V} \tilde{a}_{\Om} + e^{i\Om V} \tilde{a}^{\dag}_{\Om}
\right).
\label{free-field-mode}
\end{align}
The Unruh vacuum $|0\rangle$ is annihilated by
$a_{\Om}$ and $\tilde{a}_{\Om}$.

\subsection{An Example}
\label{Example}

As an example,
we compute in this subsection the transition amplitude
for a specific higher-derivative operator
in the background of a collapsing null thin shell.
This operator is
\begin{align}
\hat{\cal X}_n = \frac{\lam_n}{\Lam^{(n-1)(D/2+1)+m}} \,
g^{\mu_1\nu_1}\cdots g^{\mu_n\nu_n}
(\nabla_{\mu_1}\cdots\nabla_{\mu_n}\phi_1)
(\nabla_{\nu_1}\phi_2)\cdots(\nabla_{\nu_n}\phi_2) \phi_3^m.
\label{Xn}
\end{align}
For the initial and final states $|i\rangle$ \eqref{i} and $|f\rangle$ \eqref{f},
\begin{align}
\langle i |\hat{\cal X}_n| f \rangle 
\sim \frac{\lam_n}{\Lam^{(n-1)(D/2+1)+m}} \, (g^{UV})^n \int_0^{\infty} d\om_U \;
\sqrt{\frac{\om}{\om_U}} \, B^{\ast}_{\om\om_U} \,
\frac{\om_U^n e^{i\om_U U}}{r^{(m+1)(D-2)/2}} \left(\del_V \frac{1}{r^{(D-2)/2}}\right)^n,
\end{align}
where we have ignored factors of ${\cal O}(1)$.
The phase factor $\exp[i(\om' + \sum_{i=1}^{m} \om''_i)V]$ is also ignored
as we have assumed $\om', \om''_i$ to be arbitrarily close to $0$.

Define the spacetime region
\begin{align}
{\cal V} \equiv (T_0, T_1)\times (r_0, r_1)\times \Delta S^2,
\end{align}
where the time coordinate $T$ is defined as
\begin{align}
T \equiv \frac{U + V}{2},
\end{align}
and $\Delta S^2$ represents a solid angle of $\mathcal{O}(1)$.
The integral
\begin{align}
{\cal M}_{\cal X} \equiv
\int_{\cal V} d^D x \, \sqrt{-g} \, \langle i |\hat{\cal X}_n| f \rangle
\label{MX}
\end{align}
gives the transition amplitude for the process $|i\rangle \rightarrow |f\rangle$
in the spatial region $(r_0, r_1)$ over the period of time $(T_0, T_1)$
induced by the interaction ${\cal X}_n$.

The amplitude \eqref{MX} was computed for $D=4$ in Ref.\cite{Scattering}
for a range $(r_0, r_1)$ large enough to include the region inside the thin shell
as well as the region outside the thin shell.
Although the flat spacetime inside the shell has translation symmetry in time
and energy conservation forbids the creation of particles from vacuum,
the collapsing shell changes the interface
between the flat spacetime and the near-horizon region.
This has the dominant contribution to the amplitude ${\cal M}_{\cal X}$.
On the other hand, 
it was also shown that
even the contribution of the near-horizon region alone
is sufficient to induce an ${\cal O}(1)$-amplitude for particle creation
within the scrambling time \cite{Scattering}.

Apart from the numerical factor of ${\cal O}(1)$
that depends on the spacetime dimension $D$,
it is straightfoward to extend the expression of Ref.\cite{Scattering} for $D=4$
to those applicable to a generic $D \geq 4$ for $\om \sim 1/\bar{a}_{\ast}$:
\begin{align}
{\cal M}_{\cal X} \sim \frac{1}{(\bar{a}_{\ast}\Lam)^{(n-1)(D/2+1)}} \,
C^{-(n-2)}(u_s, v_s) \, \left[\frac{(T_1 - T_0)C^{-1}(u_s, v_s)}{\bar{a}_{\ast}}\right],
\label{cond-5}
\end{align}
where $v_s$ is the $v$-coordinate of the collapsing thin shell,
and $u_s$ is the $u$-coordinate of the point $(T = T_1, v = v_s)$.
\footnote{
The point $(u = u_s, v = v_s)$, or equivalently $(T = T_1, v = v_s)$,
is the point where the thin shell exits the region ${\cal V}$.
It is the point where the conformal factor $C$ takes the minimal value in ${\cal V}$
in the near-horizon region.
}
Note that $(T_1 - T_0) C^{-1}(u_s, v_s)$
is the time scale of the region ${\cal V}$ from the viewpoint
of a distant observer.

\subsection{Firewall}

According to eq.\eqref{cond-5},
for effective theories with a cutoff energy scale $\Lambda < M_p$,
a sufficient condition to satisfy the criterion \eqref{precond}
for $\hat{\cal O}_n = \hat{\cal X}_n$ is
\begin{align}
C^{n-2}(u_s, v_s) \ll 
|\lam_n| \, \frac{1}{(\bar{a}_{\ast}\Lam)^{(n-1)(D/2+1)}}.
\label{cond-6}
\end{align}
Using eqs.\eqref{C0} and \eqref{Cast} to estimate the order of magnitude of $C(u, v_s)$,
this happens around the surface of the collapsing matter when
\begin{align}
u - u_c > \frac{(D+2)(n-1)}{(D-3)(n-2)} \, \bar{a}_{\ast} \,
\log\left(\bar{a}_{\ast}\Lam\right)
\sim \mathcal{O}\left(
a\log\left(\frac{a}{\ell_p}\right)
\right),
\label{eq2}
\end{align}
where the right hand side is the scrambling time \cite{Sekino:2008he}.

The operators $\hat{\cal X}_n$ \eqref{Xn} considered above
are by no means the only higher-derivative interactions that
would lead to large transition amplitudes
and eventually the breakdown of the effective theory.
Conceivably, 
other operators of the schematic form
\begin{align}
\phi^{l} \nabla^m\phi \nabla^n\phi \cdots \nabla^k \phi
\end{align}
(where the indices on the covariant derivatives are contracted but omitted)
would have similar properties as $\hat{\cal V}_n$.
They would lead to a slightly different criterion from eq.\eqref{cond-6},
and the order-of-magnitude estimate of the time \eqref{eq2}
for an ${\cal O}(1)$-amplitude of particle creation
(the emergence of the firewall)
is expected to have a coefficient
different from $\frac{(D+2)(n-1)}{(D-3)(n-2)}$ in eq.\eqref{eq2},
but the condition for a large transition amplitude 
would still be
\begin{align}
u - u_c 
\gtrsim \mathcal{O}\left(
a\log\left(\frac{a}{\ell_p}\right)
\right).
\label{u-uc}
\end{align}

A large transition amplitude for $|i\rangle \rightarrow |f\rangle$
means that there will be a lot of particles created around the shell.
In particular, 
the state $c_{\om}|0\rangle$ is a superposition of 1-particle states
of all frequencies $\om_U$ for freely falling observers.
Although the spectrum is suppressed at high-frequency modes by the factor of $1/\sqrt{\om_U}$,
\footnote{
According to eq.\eqref{c=Aa},
the state $c_{\om}|0\rangle$ is a superposition of
1-particle states $a^{\dag}_{\om'}|0\rangle$
with the distribution function $B_{\om\om'}$.
Since $|B_{\om\om'}| \propto 1/\sqrt{\om'}$,
the spectrum of high-frequency modes is suppressed by $1/\sqrt{\om_U}$.
}
it is in fact still dominated by high-frequency modes
because of the infinite range of $\om_U$ in the limit $\om_U \rightarrow \infty$.
Hence,
the abundant production of 1-particle states around the horizon
constitute something like a ``firewall''.

We emphasize that the firewall is deduced here
as a prediction of the effective theory,
without demanding the absence of information loss.
Furthermore,
since there are infinitely many higher-derivative terms
becoming important around the same time,
the effective theory may simply break down 
before any high-energy particle flux is created.
Strictly speaking,
the emergence of the firewall is questionable,
and so is Hawking radiation.

\section{Comments}

The same estimate of the time \eqref{u-uc} for
$\mathcal{O}(1)$-correction to the horizon,
which we have derived from the condition \eqref{precond}
on the transition amplitude,
can in fact be derived from a much simpler criterion.
For a given frequency $\om$ defined with respect to $u$,
we have
\begin{align}
\om_U \equiv \left(\frac{dU}{du}\right)^{-1} \om
\label{omU-omu}
\end{align}
as the frequency defined with respect to the coordinate $U$.
Since $\left|\frac{1}{\bar{a}} \frac{d^n \bar{a}}{dV^n}\right|$
characterizes (the inverse of) the length scale of the $V$-dependence of the system,
it is natural to consider
\begin{align}
\om_U^n \, \left|\frac{1}{\bar{a}} \frac{d^n \bar{a}}{dV^n}\right| \gg \Lambda^{2n},
\label{cond-9}
\end{align}
as a heuristic generalization of the well-known criterion $\om_U \om_V \gg \Lambda^2$
for the breakdown of the effective theory.
Applying it to $\om \sim 1/a$ for the quantum fluctuation
that will turn into Hawking radiation at large distances,
we find the condition \eqref{cond-9} equivalent to
\begin{align}
C^n(u, v) \ll 
\frac{(n-1)!}{2^{n-1}} \,
\frac{\sigma \ell_p^{2n-2}}{\bar{a}^{2n-2}} \, (a\om)^n
\label{cond-10}
\end{align}
for $\Lambda \sim M_p$.
It resembles eq.\eqref{cond-6},
and leads to the same estimate of the time \eqref{u-uc} of firewall.

Since the effective theory predicts that
particles are created in the initially uneventful horizon,
their energy-momentum tensor would modify
these energy conditions \eqref{Tuu} -- \eqref{Tthth},
not long after the trapping horizon emerges.
If we wish to have an effective-theoretic description
of black holes over a time scale
longer than the scrambling time,
it is desirable to have a different, self-consistent
assumption about the vacuum energy-momentum tensor.
An example is the KMY model \cite{Kawai:2013mda}.
(See also Refs.
\cite{Kawai:2014afa,Ho:2015fja,Kawai:2015uya,Ho:2015vga,Ho:2016acf,Kawai:2017txu}.)

We made the statement, 
from the viewpoint of a distant observer,
that the exponential form of $C(u, v)$ plays an important role
in the largeness of the transition amplitude.
For freely falling observers,
the conformal factor is $1$,
and the exponential form of $C(u, v)$ is {\em a priori} irrelevant.
However,
since the transition amplitudes are invariant under coordinate transformations,
$C(u, v)$ must be simply hidden in other forms.
Indeed, $C(u, v)$ is equivalent to the product of $dU/du$ and $dV/dv$.
In Ref.\cite{Scattering},
the transition amplitude was calculated in the $(U, V)$ coordinate system,
and it was explained how the factor $C(u, v)$ appears.

A logical possibility is that,
before the firewall actually appears,
the state $c_{\om}|0\rangle$ in $|f\rangle$ \eqref{f}
cannot be suitably described within the effective theory.
This implies that what we know about Hawking radiation is unreliable.
If the state $c_{\om}|0\rangle$ is not well-defined in the effective theory,
there may be no Hawking radiation at all,
or that its temperature is much higher than $\mathcal{O}(1/a)$.
This is essentially the trans-Planckian problem \cite{trans-Planckian-1}.
Despite attempts to resolve this problem \cite{trans-Planckian-2},
there is so far no consensus on its resolution \cite{trans-Planckian-3}.
(See also discussions in Ref.\cite{Scattering}.)
This work has rigorously rephrased and 
sharpened the trans-Planckian problem.

To summarize,
this work points out a conflict between the following two features:
\begin{itemize}
\item
Uneventful horizon: \\
The energy scale of the vacuum energy-momentum tensor
is not larger than $\mathcal{O}(1/a)$
for freely falling observers comoving with the collapsing matter.
(See eqs.\eqref{Tuu} -- \eqref{Tthth}.)
\item
Hawking radiation: \\
The radiation at large distances
has a thermal spectrum at a temperature $T_H \sim \mathcal{O}(1/a)$.
This statement is based on the following assumptions:
\begin{itemize}
\item
The effective theory is valid around the horizon
and it admits a perturbative formulation.
\item
The horizon is in the vacuum state $|0\rangle$
for freely falling observers.
\item
The effective theory includes
the state $c_{\om}|0\rangle$ for $\om \sim \mathcal{O}(1/a)$,
where $c_{\om}$ is the annihilation operator for fiducial observers,
so that the spectrum $\langle 0|c^{\dagger}_{\om}c_{\om}|0\rangle$
of Hawking radiation can be computed.
\end{itemize}
\end{itemize}
Our calculation shows that
the two statements above cannot both be valid for a time longer than
the scrambling time $\mathcal{O}(a\log(a/\ell_p))$
after the collapsing matter enters the near-horizon region.

Various scenarios as alternatives to the conventional model, 
including Refs.\cite{Gerlach:1976ji,Stephens:1993an,tHooft:1996rdg,FuzzBall,Kawai:2013mda,Kawai:2014afa,Kawai:2015uya,Vachaspati:2006ki,Barcelo:2007yk,Mersini-Houghton,Baccetti:2016lsb,Mathur:2020ely},
are consistent with the conclusion above.
In these models,
information loss is not a necessity.
Although we still need to use string theory
or another theory of Planck-scale physics
to understand how information is transferred
from the collapsing matter to outgoing radiation,
we see that a careful effective-theoretic calculation
does not necessarily lead to information loss.


\section*{Acknowledgement}

We thank Heng-Yu Chen, Hsin-Chia Cheng,
Yu-tin Huang, Hsien-chung Kao,
Hikaru Kawai, Samir Mathur, Yoshinori Matsuo,
and Yuki Yokokura for valuable discussions.
This work is supported in part by the Ministry of Science and Technology, R.O.C. 
and by National Taiwan University. 

\appendix

\section{Solution to Semi-Classical Einstein Equation}
\label{Solution-EE}

Assuming the uneventful conditions,
it is straightforward to solve the semi-classical Einstein equation 
in the near-horizon region following Refs.\cite{ShortDistance,Scattering}.

In $D$-dimensional spacetime,
we find
\begin{align}
C(u, v) \simeq 
C_0(u, v) \equiv
C(u_{\ast}, v_{\ast}) \,
\frac{r^{D-3}(u_{\ast}, v_{\ast})}{r^{D-3}(u, v)} \,
e^{- \int_{u_{\ast}}^u \frac{(D-3) du'}{2a(u')}} e^{- \int_v^{v_{\ast}} \frac{(D-3) dv'}{2\bar{a}(v')}}
\label{C0-sol}
\end{align}
at the leading-order approximation,
where
\hide{
$a(u) = - \frac{D-3}{2F'}$ and $\bar{a}(v) = \frac{D-3}{2\bar{F}'}$,
respectively,
and}
$(u_{\ast}, v_{\ast})$ is an arbitrary reference point inside the near-horizon region.

Comparing this solution with the Schwarzschild solution,
we see that the two parametric functions $a(u)$, $\bar{a}(v)$
should be matched with the Schwarzschild radius.
Since Hawking temperature is of $\mathcal{O}(1/a)$,
we have
\begin{align}
\frac{da(u)}{du} \sim 
\mathcal{O}\left(\frac{\kappa}{a^{D-2}(u)}\right),
\qquad
\frac{d\bar{a}(v)}{dv} \sim \mathcal{O}\left(\frac{\kappa}{\bar{a}^{D-2}(v)}\right).
\label{dadt}
\end{align}
We will focus on a sufficiently small part of the near-horizon region
in which both $a(u)$ and $\bar{a}(v)$ are of the same order of magnitude,
which will be denoted $\mathcal{O}(a)$\footnote{
$\mathcal{O}(a) = \mathcal{O}\left((G_N M)^{1/(D-3)}\right)$
for a black hole of the initial mass $M$.
}.
According to eq.\eqref{dadt},
the ranges of the coordinates $u, v$ are restricted by
$\Delta u$, $\Delta v \ll \mathcal{O}\left(a^{D-1}/\ell_p^{D-2}\right)$
so that $\Delta a(u)$, $\Delta\bar{a}(v) \ll \mathcal{O}(a)$.

As $C(u, v) = 0$ on the horizon in the Schwarzschild solution,
we expect that,
for a reference point close to the trapping horizon,
\begin{align}
C(u_{\ast}, v_{\ast}) \sim C_0(u_{\ast}, v_{\ast}) 
\sim \mathcal{O}\left(\frac{\kappa}{a^{D-2}}\right)
\label{Cast}
\end{align}
when the quantum effect is turned on.
Since $C_0(u, v)$ \eqref{C0-sol} is furthermore
exponentially smaller deeper inside the near-horizon region,
we can solve the semi-classical Einstein equations in power expansions of $C_0$.

\hide{
Plugging the ansatz
\begin{align}
C(u, v) &\simeq C_0(u, v) + A(u, v) C^2_0(u, v) + \cdots,
\\
r(u, v) &\simeq r_0(u, v) + r_1(u, v) C_0(u, v) + \cdots
\end{align}
into the semi-classical Einstein equations,
we derive the differential equations satisfied by
the coefficients $A(u, v)$, $r_0(u, v)$ and $r_1(u, v)$.
For an adiabatic process,
the $u$, $v$ derivatives of $A(u, v)$, $r_0(u, v)$ and $r_1(u, v)$
are assumed to be higher orders in the $\kappa$-expansion.
We can then solve $A(u, v)$, $r_0(u, v)$ and $r_1(u, v)$
approximately as an expansion in powers of $\kappa$.
}

The solution as an expansion in $C_0$,
with the coefficients expanded in powers of $\kappa$ is given by
\begin{align}
C(u, v) &\simeq C_0(u, v)
- (D-2)(D-3) \frac{a(u)\bar{a}(v)}{r_0^2(v)} C_0^2(u, v)
+ \mathcal{O}\left(\frac{\kappa}{a^{D-2}} \, C_0^2\right),
\label{C}
\\
r(u, v) &\simeq r_0(v) + (D-3) \frac{a(u)\bar{a}(v)}{r_0(v)} C_0(u, v)
+ \mathcal{O}\left(\frac{\kappa}{a^{D-3}} \, C_0\right).
\label{r}
\end{align}
where $C_0(u, v)$ is given in eq.\eqref{C0-sol}.


\end{document}